\documentclass[conference]{IEEEtran}
\IEEEoverridecommandlockouts
\usepackage{cite}
\usepackage{amsmath,amssymb,amsfonts}
\usepackage{algorithm}
\usepackage{algorithmic}

\usepackage{balance}
\usepackage{booktabs}
\usepackage{courier}
\usepackage{graphicx}
\usepackage{textcomp}
\usepackage{xcolor}
\usepackage{multirow}
\usepackage{enumitem}
\usepackage{tcolorbox}

\def\BibTeX{{\rm B\kern-.05em{\sc i\kern-.025em b}\kern-.08em
    T\kern-.1667em\lower.7ex\hbox{E}\kern-.125emX}}
\begin{document}

\title{A Novel Approach for Automated Design Information Mining from Issue Logs}

\author{
    \IEEEauthorblockN{
        Jiuang Zhao\IEEEauthorrefmark{1}\IEEEauthorrefmark{2},
        Zitian Yang\IEEEauthorrefmark{1}\IEEEauthorrefmark{3},
        Li Zhang\IEEEauthorrefmark{2},
        Xiaoli Lian\IEEEauthorrefmark{2}\IEEEauthorrefmark{1}\IEEEauthorrefmark{1}, and
        Donghao Yang\IEEEauthorrefmark{2}
    }
    
    \IEEEauthorblockA{
        \IEEEauthorrefmark{2}\textit{School of Computer Science and Engineering, Beihang University, Beijing, China} \\
        \IEEEauthorrefmark{3}\textit{School of Software Engineering, Beihang University, Beijing, China}\\
        Email: \{zja1999, yangzitian, lily, lianxiaoli, yangdonghao\}@buaa.edu.cn
    }
    \thanks{\IEEEauthorrefmark{1}Co-first authors.}
    \thanks{\IEEEauthorrefmark{1}\IEEEauthorrefmark{1}Corresponding author.}
}

\maketitle

\begin{abstract}
Software architectures are usually meticulously designed to address multiple quality concerns and support long-term maintenance. However, due to the imbalance between the cost and value for developers to document design rationales (i.e., the design alternatives and the underlying arguments for making or rejecting decisions), these rationales are often obsolete or even missing. The lack of design knowledge has motivated a number of studies to extract design information from various platforms in recent years. Unfortunately, despite the wealth of discussion records related to design information provided by platforms like open-source communities, existing research often overlooks the underlying arguments behind alternatives due to challenges such as the intricate semantics of discussions and the lack of benchmarks for design rationale extraction.

In this paper, we propose a novel method, named by \textit{DRMiner}, to automatically mine latent design rationales from developers' live discussion in open-source community (i.e., issue logs in Jira).  To better identify solutions and the arguments supporting them, \textit{DRMiner} skillfully decomposes the problem into multiple text classification tasks and tackles them using prompt tuning of language models and customized text-related features. To evaluate \textit{DRMiner}, we acquire issue logs from Cassandra, Flink, and Solr repositories in Jira, and then annotate and process them under a rigorous scheme, ultimately forming a dataset for design rationale mining. Experimental results show that \textit{DRMiner} achieves an F1 score of 65\% for mining design rationales, outperforming all baselines with a 7\% improvement over GPT-4.0. Furthermore, we investigate the usefulness of the design rationales mined by \textit{DRMiner} for automated program repair (APR) and find that the design rationales significantly enhance APR, achieving 14$\times$ higher full-match repairs on average.
\end{abstract}


\section{Introduction}
\label{sec:introduction}

Software architectures must be carefully designed and implemented to address multiple quality concerns, such as performance, reliability, and maintainability \cite{Bass2012SAP}. For large-scale software systems, their architectures are intricate and they often have a long lifespan, making long-term maintenance and iterative update more critical and necessary. Moreover, without fully understanding the design rationale, i.e., the design alternatives considered and the underlying arguments for making or rejecting a decision, it is almost impossible for engineers to deal with long-term code maintenance. Although there is a plethora of content describing designs on online platforms (e.g., official forums and open-source communities), documenting the knowledge behind critical designs has not yet become a standard practice \cite{Robillard2016disseminatingarchitecturalknowledge}, resulting in these documents being missing or outdated \cite{Murphy2001SRM}\cite{Steinmacher2015SBF}. With the turnover of the original developers or the addition of new developers, the architecture inevitably becomes eroded \cite{vanGurp2002DEP}\cite{crowston2003defining}. 

When surveyed, 74\% of respondents state that they forget the reasons behind their design decisions, and 80\% say that they are not able to understand the reasons for the decisions \cite{TANG20061792}. With the evolving of software engineering, people realize that documenting \emph{how and why a design decision is made} can effectively prevent knowledge from evaporating. Nonetheless, capturing rationales takes effort that is more likely to benefit future developers and maintainers, i.e., ``downstream'' users, while requiring time and effort of ``upstream'' developers \cite{Grudin1996Evaluating}. This may reduce the incentive for decision-makers to record this information when they will not gain immediate benefit. As a result, while there are systems designed to support rationale capture and reuse\cite{Peng2009KA}\cite{jansen2005software}\cite{Tang2007RAM}\cite{Babar2007TMS}\cite{Capilla2006WTM}, these systems are not being used by practicing software developers.

In fact, software engineers are able to find design-related information from various sources, ranging from formal documentations such as design documents or requirement specifications, to informal documentations, such as discussions generated by bug reports or issue logs in open-source communities \cite{Arya2019ADITO}\cite{Rana2018HDDDR}. Fig. 1 shows a part of issue [CASSANDRA-11452] (improving one existing feature) selected from Cassandra repository in a widely used issue tracker Jira\footnote{https://www.atlassian.com/software/jira}. 
From the back and forth discussions, it is possible to identify the solutions to this issue (e.g., in the first comment) and the arguments for (e.g., in the first comment) and against (e.g., in the second comment) them. 

\begin{figure}[!htb] 	
    \centering 	
    \includegraphics[trim=53 103 150 5, clip, width=1\linewidth]{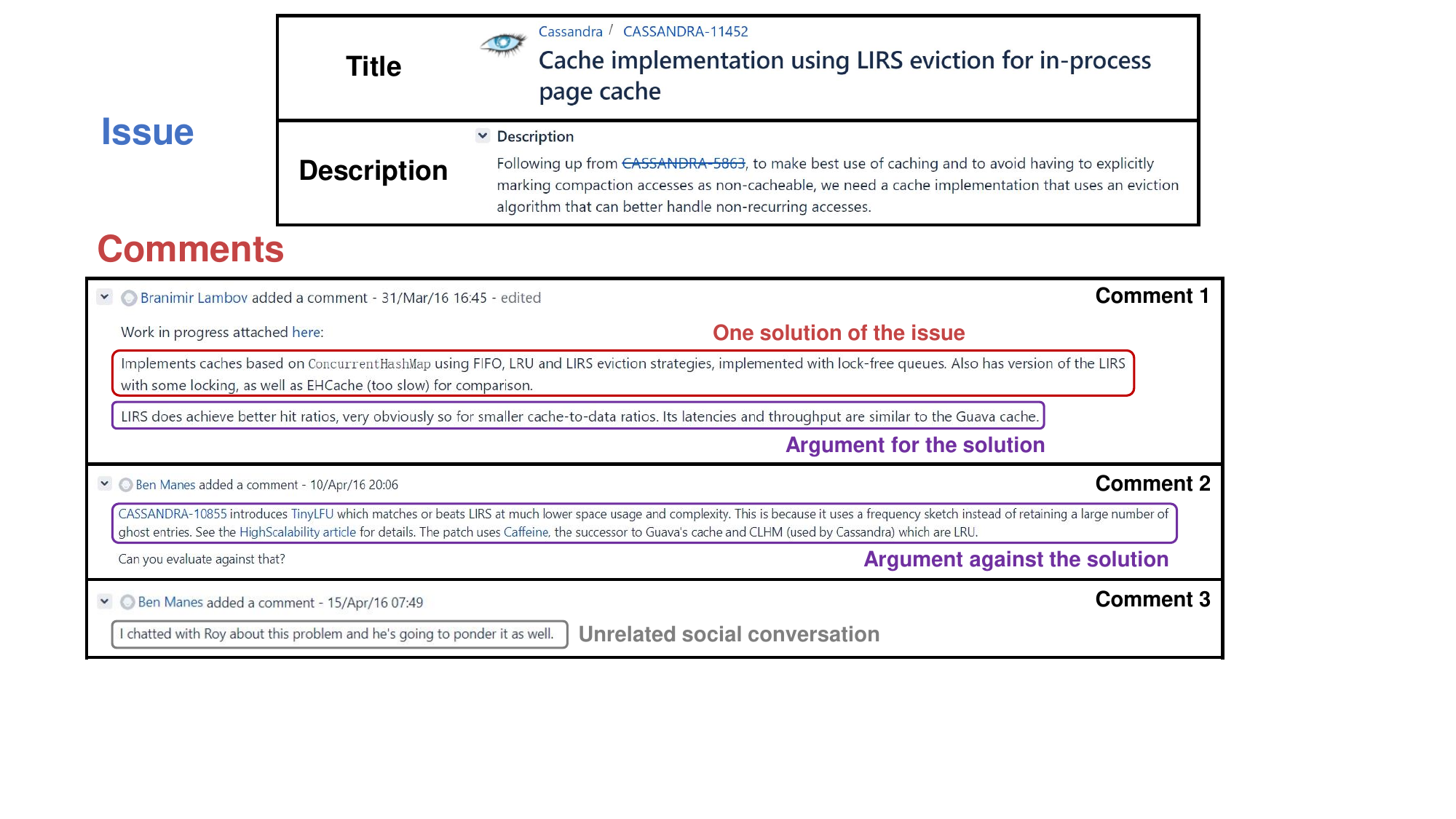} 	
    \caption{Part of CASSANDRA-11452 Selected from Jira}
    \label{fig:issuelogExample}
\end{figure}
Many studies have paid attention to uncovering these latent pieces of design information. Some of them extract designs, requirements and decisions from structured or unstructured design documents to form design rationales \cite{Lopez2012BGS}\cite{Liang2012LWD}\cite{Burge2003RationaleSF}\cite{dhaouadi2022end}. Some others mine issues, solutions and implementations from communities (e.g., from E-mails, live chats or discussion notes), and seek their potential associations \cite{Shi2021ISPY}\cite{kleebaum2021continuous}\cite{sharma2021EROSS}. However, seldom of existing researches have put effort into extracting design rationales from issue logs in open-source communities. As a primary channel for developers to post and resolve issues, open-source communities usually contain rich design knowledge that may be immensely helpful during project development and maintenance \cite{shalwani2022using}. Besides, although several existing studies have reconstructed the so-called ``design rationale'', the underlying arguments (i.e., \textit{how and why a design decision is made}) behind rationales are easily ignored \cite{sharma2021EROSS}.

Nevertheless, it is difficult to mine design rationales from issue logs in open-source communities due to the following challenges. \textbf{(1) Intricate semantics.} Discussions in issue logs appear scattered and entangled, with multiple concurrent arguments regarding different solutions frequently existing in an interleaved manner \cite{Shi2021ISPY}. The long distance between solutions and arguments places extremely high demands on method's ability to understand contextual semantics. \textbf{(2) Inconsistent definitions of design rationale.} Due to the diversity of design knowledge, there is currently no unified definition for design rationale. Some believe that design rationale includes issues, solutions, and artifacts \cite{Liang2012LWD}, while others consider rationales and decisions separately \cite{dhaouadi2022extraction}, making the objective of extraction undecidable. \textbf{(3) Lack of public benchmark.} There is presently no publicly available dataset containing design rationales within issue logs. Moreover, the process of manual annotation is time-consuming, even for experienced experts \cite{Shi2021ISPY}, which poses a significant obstacle to advancing research in this area.



In this paper, we present \underline{D}esign \underline{R}ationale \underline{Miner} (\emph{DRMiner}), a novel approach which integrates the advanced \textit{large language models (LLMs)} and the specific features of online discussions, to automatically mine design rationales from issue logs in Jira. 
\textit{DRMiner} consists of three phases, including design text extraction, design text pairing, and design rationale construction. \textit{DRMiner} first uses LLM with external sentence features to extract sentences related to rationales. Then, it pairs these sentences to obtain their relationships leveraging the instruction fine-tuning of LLM. Finally, it constructs complete design rationales based on rules.
To evaluate \textit{DRMiner}, we acquire issue logs from Cassandra, Flink, and Solr repositories in Jira, and then annotate and process the data under a rigorous scheme, finally forming a dataset for our design rationale mining task. Our experimental results demonstrate that \textit{DRMiner} outperforms all baseline methods.

Furthermore, we conduct a useful evaluation by providing the extracted rationales to advanced LLMs to explore their impact on \textit{Automated Program Repair (APR)}. The results show that the accuracy of APR significantly improves when leveraging rationales mined by \textit{DRMiner}, indicating the effectiveness of knowledge at design level in enhancing APR for specific large-scale projects. 

In summary, the contributions of this paper are:
\begin{itemize}[leftmargin=1mm]
    \item \textbf{Approach:} We introduce a novel approach named \textit{DRMiner} which enables automatically mining design rationales within issue logs in Jira, by leveraging advanced LLMs and the task specific heuristics. 
    \item \textbf{Evaluation:} Through quantitative experiments, we have demonstrated that \textit{DRMiner} significantly outperforms baseline methods in extracting design rationales. Additionally, our evaluation of its usefulness reveals substantial improvements in automated program repair when utilizing the extracted design rationales.
    \item \textbf{Benchmark:} To facilitate further empirical research, we have meticulously compiled a dataset consisting of 2092 sentences from 30 distinct issues across three large-scale projects. 
\end{itemize}

\section{Problem Statement}
\label{sec:background}



\noindent \textbf{Design Rationale.} For design rationale, many researchers have put forward their understanding \cite{gruber1991design}\cite{rogers2015using}\cite{Lester2018IDRUACO}\cite{McCall2018}. In general, design rationale is the reasons behind decisions in the life cycle of software \cite{rogers2015using} or design process \cite{Lester2018IDRUACO}. We follow this common definition.

Particularly, in this present work we aim to extract design rationale, i.e., the solutions and their related arguments, for issue resolving. The arguments for or against a solution are actually the reasons for making or rejecting this design decision.

\noindent \textbf{Issue Logs.} As shown in Fig. \ref{fig:issuelogExample}, issue logs tend to be rich in design decisions, their motivations, and implementation guidelines. The meta-data of one issue log is usually complicated. We focus on the parts beneficial for extracting design rationales, namely the \textit{issue title} which we consider as the summary of the issue, the \textit{description}, and the \emph{comments} on the design decision posted in issue summary and description, as well as the possible solutions and their corresponding supporting and opposing arguments. 

\begin{figure}[!htb] 	
    \centering 	
    \includegraphics[trim= 90 91 187 97, clip, width=1\linewidth]{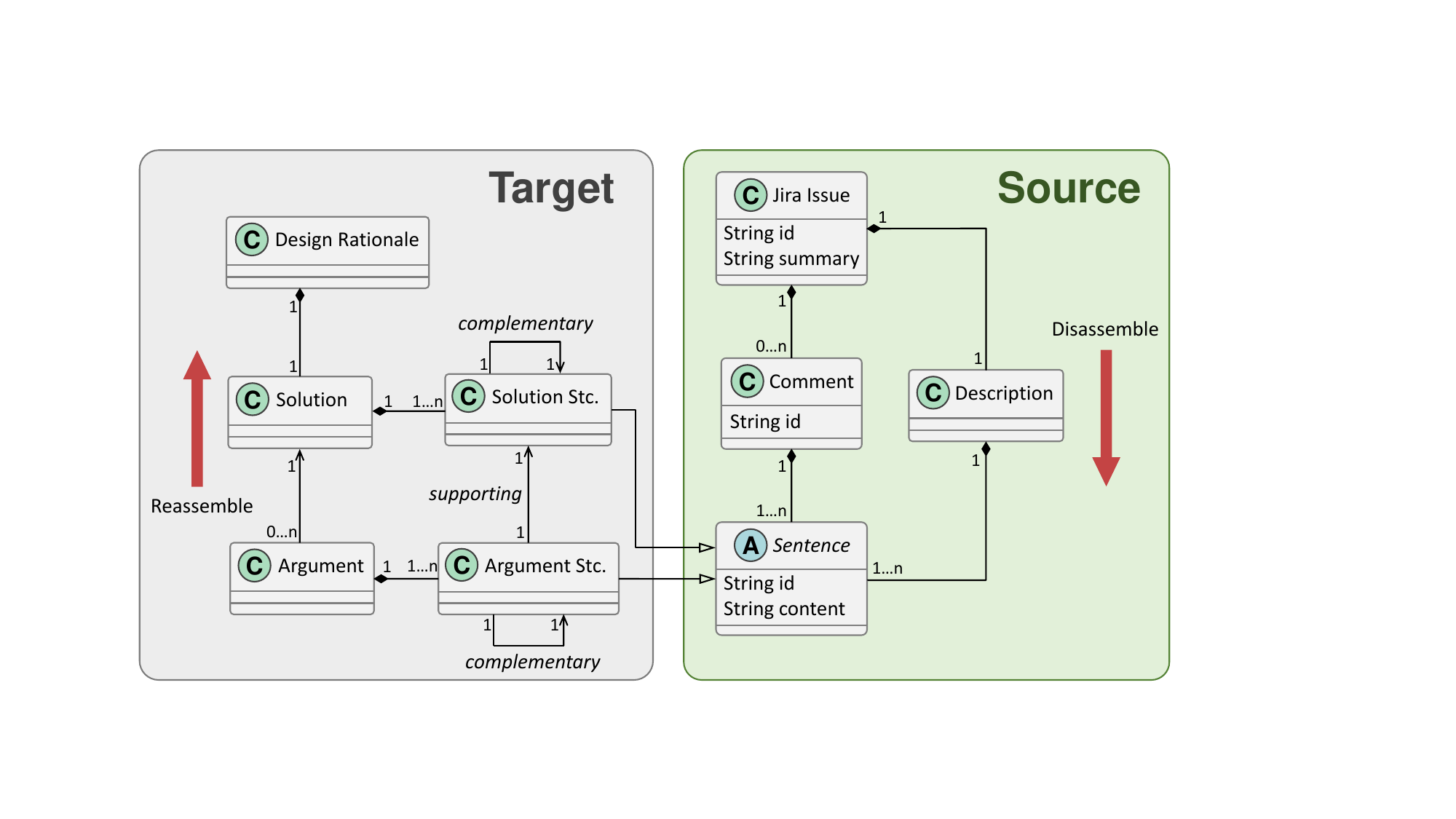} 
    \caption{Issue Log-Design Rationale Class Diagram}
    \label{fig:jiraclassdiagram}
\end{figure}

Fig. \ref{fig:jiraclassdiagram} depicts our Issue Log-Design Rationale class diagram. This diagram clarifies the relationship between Jira issue logs and the design rationale components we aim to extract, thus defining the boundaries of our research. The input for our study is the meta-data of these issue logs, with our objective being to construct design rationales for specific issues.
It is important to note that not all solutions are accompanied by explicit arguments. However, each argument does correlate with a particular solution.

The occurrence of a solution alongside its supporting arguments can vary significantly. They may be presented within a single sentence or across multiple sentences. These sentences might be contiguous or scattered throughout a comment, manifest in separate comments, or span across different types of meta-data such as issue description and comments. Therefore, our initial goal is to identify sentences relevant to both solutions and arguments, and subsequently categorize the relationships among these sentences.

Broadly speaking, there are three types of relationships between sentences. A \emph{supporting} relationship is indicated when one sentence describes a solution and another provides the corresponding arguments, whether supportive or opposing. When two sentences both describe the same solution or argument, we consider this to be a \emph{complementary} relationship. In other cases, we consider the sentences to be unrelated.



To provide a clear depiction of our problem, we present the problem formulation as follows. The design rationales we aim to extract within a single issue log $I$ are denoted by a set $\mathbb{D}_{I}=\{\mathcal{D}_i|i=1,2,\ldots,N_{D_I}\}$, where $\mathcal{D}_i$ denotes a specific design rationale, and $N_{D_I}$ is the total number of design rationales in $I$. Each $\mathcal{D}_i$ contains one Solution $\mathcal{S}_i$ and a set of Arguments $\mathbb{A}_i$, as:
\begin{equation}
    \mathcal{D}_i=\langle \mathcal{S}_i, \mathbb{A}_i\rangle
    \label{eq:DRdefine}
\end{equation}
The Solution $\mathcal{S}_i$ is composed of at least one solution sentence $S^S_{ij}$. The set $\mathbb{A}_i$ consists of supportive and opposing arguments $\mathcal{A}_{ij}$ for $\mathcal{S}_i$, where $\mathbb{A}_i=\varnothing$ is possible. Each argument $\mathcal{A}_{ij}$ is further composed of at least one argument sentence $S^A_{ijk}$. The composition of solution and arguments is shown in (\ref{eq:SAdefine}).
\begin{equation}
\begin{aligned}
    \textbf{Solution :  } & \mathcal{S}_i  =\{S^S_{ij}|j=1,2,\ldots,N_{\mathcal{S}_i}, N_{\mathcal{S}_i}\geq1\} \\
    \textbf{Arguments :  } & \mathbb{A}_i  =\{\mathcal{A}_{ij}|j=1,2,\ldots,N_{\mathbb{A}_i}, N_{\mathbb{A}_i}\geq0\} \\
    \textbf{Argument :  } & \mathcal{A}_{ij}  =\{S^A_{ijk}|k=1,2,\ldots,N_{\mathcal{A}_{ij}}, N_{\mathcal{A}_{ij}}\geq1\}
\end{aligned}
\label{eq:SAdefine}
\end{equation}

where $N_{\mathcal{S}_i}$, $N_{\mathbb{A}_i}$ and $N_{\mathcal{A}_{ij}}$ represent the total number of sentences in $\mathcal{S}_i$, arguments for $\mathcal{S}_i$, and sentences in $\mathcal{A}_{ij}$, respectively. As we formalize using sets, we do not consider the order of sentences and arguments, focusing instead on accurately extracting them. Our goal is to mine all $\mathbb{D}_{I}$ from the given issue logs.

\section{Our Approach DRMiner}
\label{sec:approach}

\begin{figure*}[!htb] 	
    \centering 	
    \includegraphics[trim= 80 62 80 54, clip, width=1\textwidth]{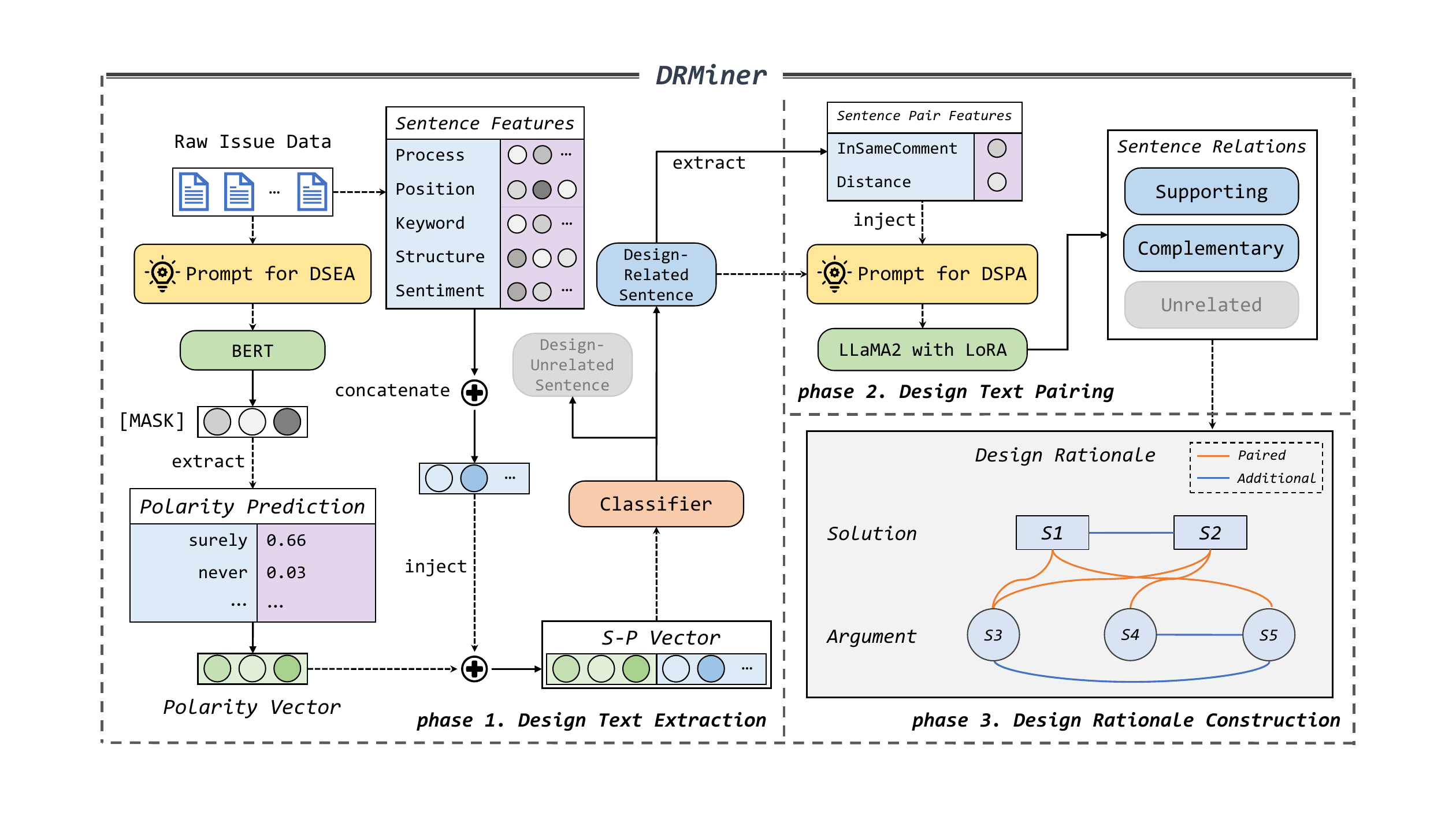} 	
    \caption{The Overall Framework of \textit{DRMiner}}
    \label{fig:drminer}
\end{figure*}

The intricate complexity of text semantics in open-source communities poses a barrier to directly mining design rationales through one-shot transfer learning. Therefore, our \textit{DRMiner} divides the task into three smaller natural language processing tasks, as illustrated in Fig. \ref{fig:drminer}: \textit{Design Text Extraction}, \textit{Design Text Pairing}, and \textit{Design Rationale Construction}.


\subsection{Design Text Extraction}
\label{sub:designtextextract}

The aim of this phase is to pinpoint design-related text, specifically, sentences that convey the design rationale information pertinent to a particular issue. 


This task is approached as a binary classification problem. We adapt \textit{Bidirectional Encoder Representations from Transformers (BERT)} as the backbone network and propose \textit{Design-related Sentence Extraction Algorithm (DSEA)}. BERT \cite{BERT}, a widely adopted pre-trained model, leverages bidirectional attention mechanism and \textit{Masked Language Modeling (MLM)} to accurately comprehend semantic information and sentence structures in text, offering advantages in our design text extraction task. 

\textit{DSEA} can be further divided into three steps. The first step is sentence polarity prediction based on prompt tuning, aiming at obtaining a polarity vector composed of the probabilities of multiple polarity words indicating the degree of the sentence is relevant to design rationales or not. This phase is accomplished by feeding the embedded prompt template into BERT. Concurrently, the second step is to identify customized sentence features from the sentence and then inject them into the polarity vector obtained from the first step, forming an S-P vector. The third step is to perform binary classification on this S-P vector using a classifier, ultimately determining whether the original sentence is design-related or design-unrelated.


\begin{figure}[!htb]
    \centering
    \includegraphics[trim= 147 57 75 22, clip,width=0.98\linewidth]{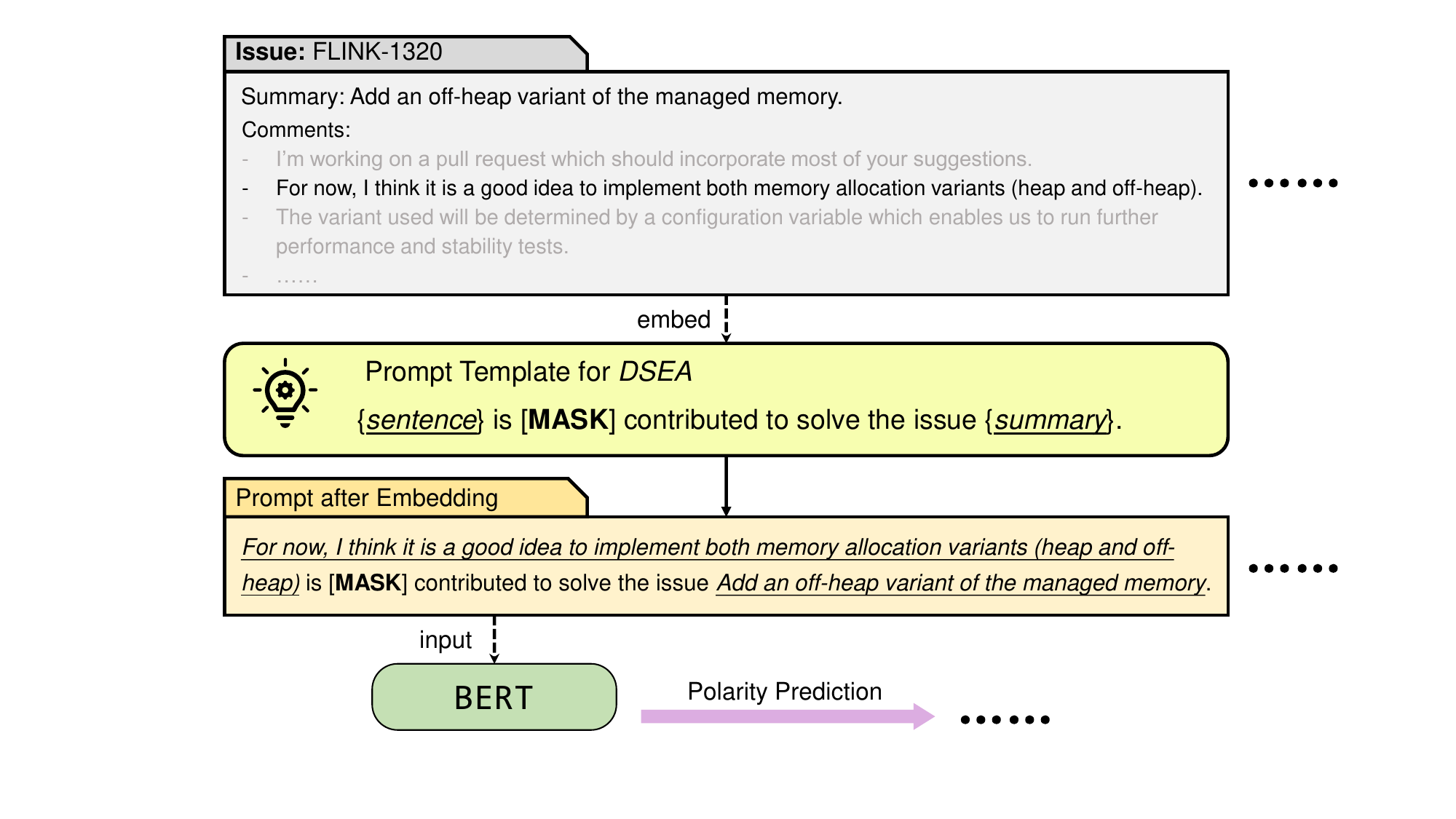}
    \caption{A Case of Sentence Polarity Prediction Using Prompt Template in \textit{DSEA}}
    \label{fig:dseapt}
\end{figure}

\textbf{Sentence Polarity Prediction Based on Prompt Tuning.} \textit{Prompt tuning} is an emerging technique in natural language processing by selecting the appropriate prompts to manipulate the model behavior so that the pre-trained language models itself can be used to predict the desired output \cite{liu2023prompt}. For design text extraction, \textit{DSEA} utilize a prompt to guide BERT in determining whether a sentence is the design text for a specific issue. A case of sentence polarity prediction using prompt template in \textit{DSEA} is illustrated in Fig. \ref{fig:dseapt}, taking an issue from Jira with key FLINK-1320 for example. 

Specifically, \textit{DSEA} use a \textit{cloze prompt} \cite{liu2023prompt} which fill in the
blanks of a textual string. There are totally three blanks that need to be filled. The \{\textit{\underline{sentence}}\} and \{\textit{\underline{summary}}\} blanks at the beginning and end of the prompt are sentences embedded by \textit{DSEA} before trainning or testing. The \{\textit{\underline{sentence}}\} placeholder signifies each sentence within the metadata of an issue, which needs to be evaluated to determine if it is design-related text (including \{\textit{\underline{summary}}\}), while the \{\textit{\underline{summary}}\} placeholder corresponds to the summary of the issue.
 The fixed description in between with the third blank [\textbf{MASK}] is used to indicate the relationship between the above two blanks, where BERT need to predict the word at [\textbf{MASK}]. 

Due to BERT has a limited sequence length, it is important to perform a pre-truncation to ensure that the word at [\textbf{MASK}] will not be truncated by BERT. We only truncate the \{\textit{\underline{sentence}}\} back to front as we can not truncate the fix description and the \{\textit{\underline{summary}}\}. The maximum length of \{\textit{\underline{sentence}}\} in \textit{DSEA} ( $\mathcal{L}_{sentence\_DSEA}$) is shown in (\ref{eq:seqlengthdsea}): 

\begin{equation}
    \mathcal{L}_{sentence\_DSEA}=\mathcal{L}_{BERT}-\mathcal{L}_{summary}-8
    \label{eq:seqlengthdsea}
\end{equation}

where $\mathcal{L}_{BERT}$ is the maximum sequence length BERT can process, and $\mathcal{L}_{summary}$ is the length of \{\textit{\underline{summary}}\}. The subtraction of 8 is due to the length of the fixed description being 7, and a special token \texttt{[CLS]} before every sentence added by BERT. If the length of \{\textit{\underline{sentence}}\} exceeds this maximum length, it will be truncated to $\mathcal{L}_{sentence\_DSEA}$.

To enhance model performance, we avoid directly prompting BERT to classify a sentence's relevance to an issue. Prior research \cite{gao2021making} indicates that aligning the label words (i.e., [\textbf{MASK}] in DSEA's prompt) with corresponding ``semantic classes'' improves accuracy. Consequently, rather than constructing a straightforward prompt such as ``{sentence} is [\textbf{related/unrelated}] to the issue {summary},'' we opt for a prompt where the [\textbf{MASK}] is filled with \emph{polarity indicators} that suggest the degree of relevance.

To cover a broader range of semantics expressible by BERT and enhance the fault tolerance of the prompt to ensure model stability, we summarize a series of commonly used adverbs expressing sentiment polarity. These adverbs are categorized into those representing positivity including [surely, directly, closely, certainly, strongly, highly, absolutely] and those representing negation including [not, un, never, no, little, hardly, rarely]. It is worth noting that we have confirmed their presence in BERT vocabulary so that the probability of these polarity words can always be extracted to support the subsequent classification. Finally, \textit{DSEA} extracts the probability of all these polarity words from the probability distribution vector (so-called representation vector) of the word at [\textbf{MASK}] output by BERT, and concatenates them to form a polarity vector. 

In summary, this initial step involves utilizing BERT's nuanced semantic understanding by feeding it prompts in a manner reminiscent of its original Masked Language Modeling (MLM) training. This process enables BERT to predict the polarity of each sentence within the issue metadata without any further training.

\textbf{Customized Sentence Features Extraction}. In text classification tasks, BERT can deeply understand the context and meaning of each sentence, capturing subtle linguistic nuances through these high-dimensional features. Nonetheless, by manually analyzing and extracting the features of sentences in certain dimensions, we can provide the model with a more comprehensive perspective to understand the text, thereby enhancing the accuracy and robustness of the extracted design text. We divide these features into five dimensions, including \textit{issue process}, \textit{position}, \textit{keyword}, \textit{text structure}, and \textit{sentiment}. 

\textbf{1) Issue Process Dimension}. Process-related features pertaining to an issue track its progression from initiation to resolution within the Jira framework. We propose features based on the following observations: ({\romannumeral1}) An issue description is more likely to propose a fresh design, thereby triggering developers to discuss it. ({\romannumeral2}) Individuals participating in the discussion may play various roles. In general, initiators of the issue and developers with relevant experience tend to contribute more comments than those who happened to come across the issue. ({\romannumeral3}) More active discussions on an issue may imply more design information. Therefore, we propose five process-related features: \textit{IsDes}, \textit{IsCreator}, \textit{AuthorCommentsCount}, \textit{CommentsCount}, and \textit{SentencesCount}. These features are unrelated to the sentence itself, only related to the discussion process in the issue.

\textbf{2) Position Dimension}. During data preparation, we found that most individuals tend to propose their alternative solution at the beginning or end of the comment. Moreover, comments appearing earlier in the discussion tend to contain more design information. Based on this observation, we propose three position-related features: \textit{CommentIndex}, \textit{SentenceIndex}, and \textit{GlobalIndex}. For \textit{CommentIndex} and \textit{SentenceIndex}, we focus on the relative positions, while for \textit{GlobalIndex}, we pay more attention to the absolute positions, as both positions may be indicative.

\textbf{3) Keyword Dimension}. Certain words can signal the introduction of a solution, such as `therefore', while others may point to an ongoing argument, like `however'. Consequently, we have compiled a list of 14 keywords to serve as features.  These include the interrogatives and relative pronouns known as the 5W1H: [what, why, when, who, which, how]; three types of modal auxiliary verbs: [should,shall,can,could,may,might]; two punctuation marks: [?,!]; greeting words [hi,hello,bye,thanks,thx,thank], causal words [so,therefore,then], and transitional words [but,yet, however]. The detection of these keywords is based on a complete matching pattern with case-insensitive; if a keyword is only a sub-string of a complete word, it will not be recognized.

\textbf{4) Text Structure Dimension}. Sentences that include code snippets or URLs tend to offer practical solutions or ideas. Furthermore, a higher word count in a sentence often suggests a richer trove of information. In light of this, we have delineated text structure features which encompass three aspects: \textit{HasCode} (a boolean indicating the presence of code), \textit{HasURL} (a boolean indicating the presence of URLs), and \textit{WordsCount} (an integer representing the number of words in a sentence).

\textbf{5) Sentiment Dimension.} The inherent polarity of a sentence is a critical indicator of the presence of supporting or opposing arguments. We employ sentiment feature extraction using the Valence Aware Dictionary and sEntiment Reasoner (VADER) \cite{hutto2014vader}, which is finely tuned for assessing sentiments conveyed in social media contexts. Our design for sentiment features draws upon the scores generated by VADER: \textit{PosSentiment}, \textit{NeuSentiment}, \textit{NegSentiment}, and \textit{CompoundSentiment}. The first three features quantify the respective proportions of content within a sentence that express positive, neutral, and negative sentiments. Conversely, \textit{CompoundSentiment} aggregates these aspects to provide an overall sentiment score through composite analysis.

\textbf{Binary Classification.} \textit{DSEA} merges the vector representing sentence features with the polarity words vector to construct an integrated one known as the \textit{S-P vector}. Subsequently, a fully connected layer is appended to BERT's output layer, serving as a classifier. This classifier is responsible for discerning whether a sentence bears relevance to design aspects or if it is unrelated to design considerations.

\subsection{Design Text Pairing}
\label{sub:designtextpairing}

The objective of this task is to predict the relationships between each pair of the extracted design-related sentences; specifically, whether they are \emph{supporting}, \emph{complementary}, or \emph{unrelated}, as discussed in Section \ref{sec:background}. We introduce the \textit{Design-related Sentence Pairing Approach (DSPA)}, which frames the pairing of design-related text as a ternary classification problem, aiming to infer the relationship between any two given sentences within the design discourse.

\textit{DSPA} is a combination of two steps. The first step is prompt construction with customized sentence pair features. The objective of this step is to form a complete prompt with instruction, two sentences and their correlated features, which will be used in the subsequent model fine-tuning. The second step is ternary classification based on the \textit{Instruction Fine-Tuning (IFT)} \cite{zhang2023llama} of LLaMA2. IFT is another form of prompt tuning \cite{liu2023prompt}. It provides guidance to language models in a formalized way (i.e., the prompt constructed in the first step) for correctly generating the expected response. 

In the second step, We adapt LLaMA2 \cite{touvron2023llama} as the backbone model, since it demonstrates deeper semantic understanding in scenarios like dialogue understanding \cite{labruna2024dynamic}. 
However, full-scale IFT of LLaMA2 poses challenges such as high resource requirements and complex hyper-parameters. Hence, we employ \textit{Low-Rank Adaptation (LoRA)} \cite{hu2021lora}, which reduces the additional parameters by introducing two low-rank matrices while keeping the original parameters unchanged, significantly reducing resource consumption and making IFT flexible and controllable. 


\begin{figure}[htb!]
    \centering
    \includegraphics[trim= 145 0 70 0, clip, width=1\linewidth]{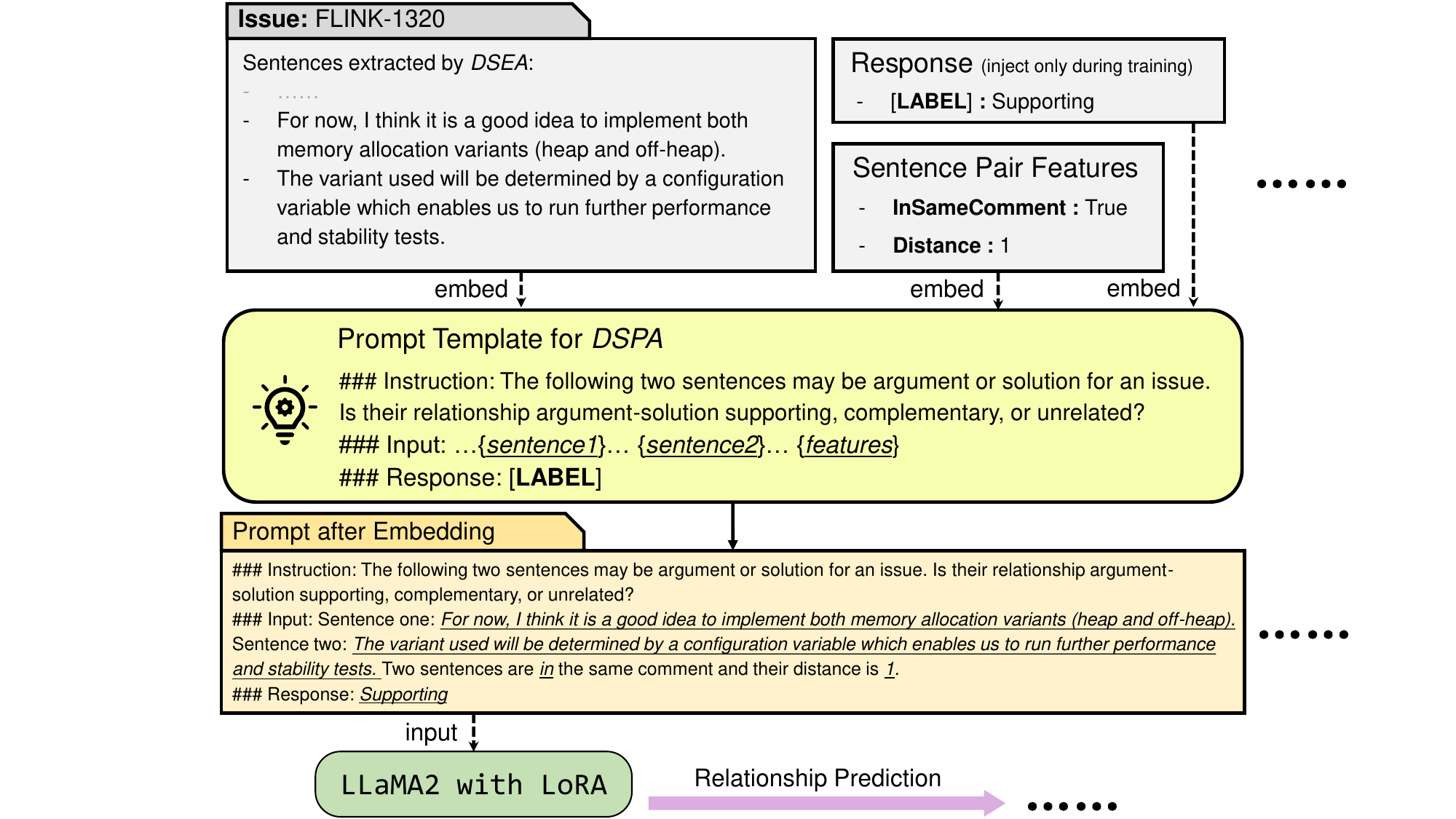}
    \caption{A Case of Sentence Pair Relationship Prediction Using Prompt Template in \textit{DSPA}}
    \label{fig:dspapt}
\end{figure}

\textbf{Prompt Construction}. Following the IFT methodology and the \textit{cloze prompt} manner, \textit{DSPA} divides the prompt template into three sections: instruction, input, and response, separated by hashtags. A case of sentence pair relationship prediction using prompt template in \textit{DSPA} is illustrated in Fig. \ref{fig:dspapt}, taking an issue with key FLINK-1320 for example. The detail of template components is described as follows.

\begin{itemize}[leftmargin=2mm]
    \item \textbf{Instruction}. \textit{Instruction} is a fixed description that depicts the background of design text pairing. In \textit{DSPA}, it is described as follow: \textit{The following two sentences may be argument or solution for an issue. Is their relationship argument-solution supporting, complementary, or unrelated?} By saying this, \textit{DSPA} restricts the response of the model to be one of the three relationship labels discussed in Section \ref{sec:background}.
    \item \textbf{Input}. The \emph{input} comprises two sentences undergoing analysis (i.e., two of the sentences \textit{DSEA} extracted), accompanied by customized features designed to augment LLaMA2's performance in classifying relationships. The definitions of these features will be elucidated in the subsequent section.
    \item \textbf{Response}. \textit{Response} represents the relationship labels of sentence pairs inferred by \textit{DSPA}, i.e., \emph{supporting}, \emph{complementary} and \emph{unrelated}. Note that these results of sentence pairing will be recorded orderly. By adding "argument-solution" in instruction, for pairs with a \textit{supporting} relationship, the first sentence will always be the argument and the second sentence will always be the solution.
    
    
    
    
\end{itemize}

Due to LLaMA2 has a limited sequence length, it is important to perform a pre-truncation to ensure that
critical information, particularly pivotal prompting terms \texttt{[Instruction]}, \texttt{[Input]}, and \texttt{[Response]}, remains intact. Similarly, we truncate \{\textit{\underline{sentence1 }}\} and \{\textit{\underline{sentence2 }}\} back to front. The maximum length of them in \textit{DSPA} ($\mathcal{L}_{sentence\_{DSPA}}$) is shown in (\ref{eq:seqlengthdspa}): 

\begin{equation}
    \mathcal{L}_{sentence\_DSPA}=\lfloor \frac{\mathcal{L}_{LLaMA2}-\mathcal{L}_{template}}{2}\rfloor
    \label{eq:seqlengthdspa}
\end{equation}

where $\mathcal{L}_{LLaMA2}$ represents the maximum sequence length that LLaMA2 can process, and $\mathcal{L}_{template}$ denotes the cumulative length of the fixed part in the template—this encompasses the instruction, response, and input but excludes the two sentences intended for embedding. We consider a pair of sentences as a matched set, hence the available sequence margin is halved to ensure equitable treatment between them. Should either sentence exceed 
$\mathcal{L}_{sentence\_DSPA}$ in length, truncation will be applied to the surplus sequence.
  



\textit{Customized Sentence Pair Features}. To improve the performance of pairing solutions with their corresponding arguments, we observed the prevalence of solutions and their discussions and identified two types of positional cues: 

({\romannumeral1}) In issue logs, developers tend to articulate comprehensive opinions within a single comment. As a result, sentences originating from separate comments are more apt to be unrelated.

({\romannumeral2}) The greater the distance between two sentences within the text, the more likely they are to be unrelated, which is a supplement to the first cue.
To leverage these insights, we extract two heuristic features for pairs of sentences:
\begin{itemize}[leftmargin=2mm]
\item \emph{InSameComment (type: boolean)} indicates whether the two sentences are from the same comment.
\item \emph{Distance (type: integer)} represents the absolute value of the difference in positions between two sentences, e.g., the \emph{Distance} between two consecutive sentences is 1.

\end{itemize}

When incorporating these sentence pair features into the prompt template, we append ``Two sentences are [\textit{\underline{InSameComment}}] the same comment and their distance is [\textit{\underline{Distance}}]'' at the end of the Input section. For specific instances, [\textit{\underline{InSameComment}}] is substituted with either `in' or `not in,' and [\textit{\underline{Distance}}] is replaced with the actual numerical distance separating the sentences.

\textbf{Ternary Classification Based on IFT.} In this step, the prompt with IFT format obtained in the first step is input into LLaMA2 with LoRA. Then, the model predicts the word at [\textbf{LABEL}] in Response section of the prompt as the ternary classification result of \textit{DSPA}. Due to the instruction, the [\textbf{LABEL}] word is restricted to one of \textit{supporting}, \textit{complementary}, or \textit{unrelated}.

\noindent $\triangleright$ \textbf{Design Rationale Construction}

So far, we have obtained all the design-related sentences in issue $I$ together with the relationships between them. The only gap between this and the ultimate design rationales $\mathbb{D}_{I}$ we aim to construct is that we have not yet separated the solutions and arguments belonging to different design rationales. Fortunately, utilizing the results of \textit{DSEA} and \textit{DSPA}, we can easily construct this set of design rationales following a series of intuitive rules, i.e., the third phase of \textit{DRMiner}.

To be specific, we firstly identify all solution sentences $\cup S^S_{ij}$ based on the \textit{supporting} relationships. Next, according to the one-to-one relationship between solutions and design rationales illustrated in Section \ref{sec:background}, we separate and classify solution sentences into corresponding $\mathcal{S}_i$ based on the \textit{complementary} relationships among solution sentences to create $\mathcal{D}_i$. Then, we assign the argument sentences to their corresponding $\mathcal{S}_i$ based on the \textit{supporting} relationships, and partition them into different $\mathcal{A}_{ij}$ according to the \textit{complementary} relationships between argument sentences to form $\mathbb{A}_i$. Ultimately, \textit{DRMiner} obtains the set $\mathbb{D}_I$ containing all design rationales in the issue $I$, along with their solutions and arguments. 
\section{Experimental evaluation}
\label{sec:experimental}
\subsection{Research Questions}
\label{subsec:rqs}
We evaluate \textit{DRMiner} to answer the following three questions:
\begin{itemize}[leftmargin=3mm]
    \item \textbf{RQ1}: How effective is \textit{DRMiner} at extracting design rationales?
    \item \textbf{RQ2}: How effective are customized sentence features used in \textit{DSEA}?
    \item \textbf{RQ3}: Can design rationales mined by \textit{DRMiner} benefit automated program repair?
\end{itemize}

The first research question aims to evaluate the overall performance of \textit{DRMiner} from design rationale level and sentence level. At design rationale level, we focus on whether the extracted design rationales are correct. At sentence level, we focus on the accuracy of the solution and argument sentences within extracted rationales. 


For RQ2, we aim to understand the importance of introducing customized design knowledge in design text extraction and the contribution of each dimension to \textit{DSEA}. This helps us discover more useful dimensions for design text extraction, enabling us to increase their weight or expand their scope.

Furthermore, to explore the impact of the design rationales mined by \textit{DRMiner} on APR, i.e., their potential to enhance the repair performance of advanced LLMs, we formulated RQ3 to conduct a useful evaluation and compare the number of fully correct program repairs.

\subsection{Data Preparation}
\label{subsec:data}

In order to automate the mining of design rationales, it is necessary to identify their existence in open-source community and construct a dataset. We select issues from three projects tracked by Jira: \textit{Flink}, \textit{Cassandra}, and \textit{Solr}, as the source of our dataset. Flink\cite{Flink} is a unified stream processing and batch processing framework; Cassandra\cite{Cassandra} is a high performance distributed NoSQL database; Solr\cite{Solr} is an enterprise search platform based on Apache Lucene\cite{Lucene}. We select 10 typical issues from each project, covering various software architecture-related designs such as caching, scheduling, and authentication.

During the pre-processing of raw data, the following texts are  filtered out: ({\romannumeral1}) automatic replies from Jira bots and GitHub bots in comment area. ({\romannumeral2}) references to previous comments. ({\romannumeral3}) special characters such as emojis that can not be encoded properly. ({\romannumeral4}) code snippets and URLs, as they are not the focus of our attention. We replace them with \texttt{[code]} and \texttt{[URL]} respectively to maintain the original text structure.

Due to the subjective nature of annotation, we design a rigorous data annotation scheme. For annotators, We recruit 16 participants with backgrounds in software engineering, including two researchers with over 12 years of academic experience, one of whom is a professor in University of Notre Dame and the other is an associate professor in Colorado College; seven doctoral students and six junior students majoring in software engineering from Beihang University, two of whom are international students, with diverse research interests on software architecture, requirement mining, and open-source community; one software developer in ByteDance with over 5 years of engineering experience. 

Subsequently, we allocate the same issue data to two different annotators and ask them to do the annotation independently. We provide each annotator the summary of our research, the guidelines for annotation, one contrived example from an imaginary issue in a system for monitoring unmanned aerial vehicles (UAV), and one real but complicated example selected from Cassandra repository. Both two examples are annotated by our researchers. After receiving two independent annotations, we conduct a discussion among the two original annotators and a third expert. For any discrepancies, we follow a majority vote to determine the final result. The final scale of the dataset is illustrated in Table \ref{tab:dataset}.

\begin{table}[!htbp]
\caption{The scale of dataset for \textit{DRMiner} after annotation}
\begin{center}
\begin{tabular}{c|c|c|c|c}
\toprule[1pt]
\multirow{2}{*}{\textbf{Project}}  & \multicolumn{3}{c|}{\textbf{Sentence Type}} & \multirow{2}{*}{\textbf{Total}} \\
\cmidrule{2-4}
 & \textbf{Solution} & \textbf{Argument} & \textbf{Unrelated} &  \\
 \midrule
 Flink      & 111 & 78  & 232 & 421 \\
Solr        & 172 & 211 & 609 & 992 \\
Cassandra   & 60  & 136 & 483 & 679 \\
\midrule
\textbf{Total} & 343 & 425 & 1324 & 2092 \\
\bottomrule[1pt]

\end{tabular}
\label{tab:dataset}
\end{center}
\end{table}
\textit{DRMiner} mines design rationales on a per-issue basis in practical, so we partition the dataset accordingly. We randomly select one issue from each of the three projects to add to the test set, while the remaining 27 issues are used for the training set.

\subsection{Environment and Implementation Detail}
\label{subsec:expesetting}

We implement \textit{DRMiner} in Python using Pytorch. The experiments are run on a computer with GeForce RTX 4090 GPU, Intel Core i7-12700KF@3.61GHz CPU, 32GB RAM, and Ubuntu 20.04 LTS Operating System.

In \textit{DSEA}, we choose the \textit{bert-large-uncased} with 340 million parameters as the backbone model. To accommodate the customized features injection in Section \ref{sub:designtextextract}, we add a $43\times2$ fully connected layer after the output layer of BERT. To train BERT, we employ Cross-Entropy Loss and AdamW optimizer with a weight\_decay of 0.01. The learning rate is set to $1e^{-5}$ and the epoch is set to 30. The maximum length that BERT can process is defined as 384, and for those shorter than 384, we use \texttt{[PAD]} to pad them.

In \textit{DSPA}, we choose the \textit{LLaMA2-7B} with 7 billion parameters as the backbone model. Three parameters are required in LoRA\cite{hu2021lora}: \textit{r}, \textit{lora\_alpha}, and \textit{lora\_dropout}. \textit{r} determines the rank used by LoRA; \textit{lora\_alpha} controls the extent to which the low-rank approximation affects the original weight matrices; \textit{lora\_dropout} behaves similarly to traditional dropout \cite{srivastava2014dropout}. We set the value of \textit{r} to 8, \textit{lora\_alpha} to 16, and \textit{lora\_dropout} to 0.05. The learning rate is set to $1e^{-4}$ and the epoch is set to 5. The optimizer and loss are the same as those used for \textit{DSEA}. The maximum length that LLaMA2 can process is defined as 512, and we do not pad those shorter than 512.


\subsection{Baselines}
\label{subsec:baseline}

The key subtasks of \textit{DRMiner} are text classification tasks. Therefore, from a feasibility perspective, many learning-based methods can well accomplish the design rationale mining task. We compare \textit{DRMiner} with three types of representative and publicly available learning-based approaches, including \textit{Machine Learning (ML)-based}, \textit{Deep Learning (DL)-based}, and \textit{Large Language Model (LLM)-based} methods as baselines, described as follows:

\begin{itemize}[leftmargin=1mm]
 \item \textbf{ML-based} methods include 5 typical classifiers, specifically Random Forest \cite{Randomforests}, Decision Tree \cite{Decisiontree}, Gradient Boosting \cite{gradientboosting}, AdaBoost \cite{adaboost}, and XGBoost \cite{xgboost}. We use \textit{Term Frequency-Inverse Document Frequency (TF-IDF)}  \cite{TFIDF} as features to classify the text.
    
 \item \textbf{DL-based} methods include 4 advanced methods on design knowledge mining. \textit{DECA} \cite{DECA} is a semi-supervised method to classify the content of development emails according to their purpose. \textit{CNC} \cite{CNC} is a novel approach that employs \textit{Convolutional Neural Networks (CNN)} to classify comments in online issue reports. Pranjal et al.\cite{srivastava2022argument} employs attention-based embedding for argument mining in online discourses. Zeberga et al.\cite{zeberga2022novel} proposes a method based on BERT and Bi-LSTM to mine depression-related posts on social media.
    
 \item \textbf{LLM-based} method. We choose the state-of-the-art LLM \textit{GPT-4.0-turbo} \cite{achiam2023gpt} due to its outstanding performance in a wide range of NLP tasks like sentence analysis and information extraction.
\end{itemize}
Due to the complexity of our goal, it is hard to directly transfer these baselines to design rationale mining. Thus, for all baselines, we follow the three-phase framework of \textit{DRMiner} to implement them. We adjust their interfaces to fit the design text extraction and design text pairing tasks respectively, and then integrate them as baseline methods for design rationale mining. We set all the hyper-parameters for baselines as specified in the original papers.

\subsection{RQ1: How effective is \textit{DRMiner}?}
\label{sec:rq1}

\begin{table*}[!htbp]
\caption{The comparison of DRMiner and baselines in design rationale level and sentence level.}
\begin{center}
\begin{tabular}{ccccccccccc}
\toprule[1pt]
\multirow{2}{*}{\textbf{Type}} & \multirow{2}{*}{\textbf{Method}} & \multicolumn{3}{c}{\textbf{Design Rationale}} & \multicolumn{3}{c}{\textbf{Solution Sentence}} & \multicolumn{3}{c}{\textbf{Argument Sentence}} \\
\cmidrule{3-5}  \cmidrule{6-8} \cmidrule{9-11}
 &  & \textit{precision} & \textit{recall} & \textit{F1-score} &   \textit{precision} & \textit{recall} & \textit{F1-score} & \textit{precision} & \textit{recall} & \textit{F1-score} \\
 \midrule

\multirow{5}{*}{ML-based} 
& Random Forest & 0 & 0 & 0 & 0 & 0 & 0 & 0 & 0 & 0\\
& Decision Tree & 0.29 & 0.13 & 0.17 & 0.11 & 0.14 & 0.12 & 0.19 & 0.13 & 0.15\\
& AdaBoost & 0.07 & 0.06 & 0.06 & 0.06 & 0.03 & 0.04 & 0 & 0 & 0\\
& Gradient Boosting & 0.18 & 0.13 & 0.15 & 0.11 & 0.07 & 0.08 & 0 & 0 & 0\\
& XGBoost & 0.15 & 0.13 & 0.14 & 0.14 & 0.07 & 0.09 & 0 & 0 & 0\\
\midrule
\multirow{4}{*}{DL-based} 
& CNC\cite{CNC} & 0.14 & 0.12 & 0.12 & 0.08 & 0.05 & 0.06 & 0 & 0 & 0\\
& DECA\cite{DECA}& 0.21 & 0.14 & 0.17 & 0.17 & 0.11 & 0.13 & 0 & 0 & 0\\
& Pranjal et al.\cite{srivastava2022argument} & 0.29 & 0.56 & 0.38 & \textbf{0.29} & 0.48 & 0.36 & 0.50 & 0.10 & 0.17 \\
& Zeberga et al.\cite{zeberga2022novel} & 0.35 & 0.63 & 0.45 & 0.26 & 0.48 & 0.34 & \textbf{0.60} & 0.10 & 0.14 \\
\midrule
LLM-based & GPT-4.0 & \textbf{1.00} & 0.69 & 0.61 & 0.24 & 0.72 & 0.37 & 0.16 & 0.44 & 0.24 \\
\midrule
Ours & DRMiner & 0.50 & \textbf{0.94} & \textbf{0.65} & \textbf{0.29} & \textbf{0.83} & \textbf{0.42} & 0.40 & \textbf{0.87} & \textbf{0.55} \\

\bottomrule[1pt]

\end{tabular}
\label{tab:rq1result}
\end{center}
\end{table*}

We use three commonly-used metrics to evaluate \textit{DRMiner}, i.e., \textit{precision}, \textit{recall}, and \textit{F1-score}. However, since design rationales encompass numerous solution and argument sentences, evaluating the correctness of a design rationale is inherently complex. To address this problem, we redefine the concepts of true positive (TP) and false positive (FP) for design rationale as follows. (\romannumeral 1) \textit{TP}: A design rationale mined by \textit{DRMiner} is considered a TP if at least one solution sentence in this rationale matches a solution sentence in the ground truth design rationales. (\romannumeral 2) \textit{FP}: A design rationale mined by \textit{DRMiner} is considered an FP if none of the solution sentences in this rationale match any solution sentence in the ground truth design rationales. For \textit{F1-score}, we follow its conventional definition as the harmonic mean of \textit{precision} and \textit{recall}. When comparing the performances, we care more about F1 since it is balanced for evaluation. 

The left half of Table \ref{tab:rq1result} shows the results of the design rationales mined by \textit{DRMiner} and baselines. We find that \textit{DRMiner} achieves the best performance among all baselines in design rationale level, with an improvement of 6.6\% compared with GPT-4.0. We have two further observations. 

(1) Compared to methods utilizing Transformer in backbone model, those not leveraging Transformer (i.e., all ML-based methods, CNC, and DECA) exhibit significantly poorer performance, with some failing to identify any valid design rationales. The absence of Transformer’s robust contextual understanding capabilities renders these methods less effective at classifying text based on local word features, especially when dealing with the complex semantics found in issue logs. 

(2) Although \textit{DRMiner} and GPT-4.0 have similar F1 scores, their performance in precision and recall is markedly different. GPT-4.0 excels at extracting fewer but more precise design rationales (with precision being 100\%), while \textit{DRMiner} tends to uncover a larger number of design rationales (with recall being 94\%), albeit with less accuracy. Nevertheless, the challenge of identifying design rationales from vast issue logs is significantly greater than verifying the correctness of extracted rationales. Therefore, we consider our \textit{DRMiner} to be more effective and practical compared to GPT-4.0.

Furthermore, as discussed in Section \ref{sec:introduction}, the arguments supporting the solutions are also of particular interest to us. Due to the relatively weak criteria for determining whether a design rationale is a true positive, in order to investigate the completeness of the mined design rationales, we compute the \textit{precision}, \textit{recall}, and \textit{F1-score} for solution and argument sentences at the sentence level, as shown in the right half of Table \ref{tab:rq1result}. Here, a sentence is considered a true positive only if its category is correct (i.e., a solution sentence or an argument sentence in a design rationale) and corresponds to the correct design rationale.

We attribute the superiority of \textit{DRMiner} over other methods primarily to our adoption of prompt tuning and customized sentence-related features. In \textit{DSEA}, we pay attention to the consistency between BERT's MLM pre-training mechanism and the cloze prompt manner, and adjust the prompt format to broaden the scope of effective words at  [\textbf{MASK}], thereby enhancing the performance and stability of text classification. In \textit{DSPA}, the IFT approach we employ accurately and concisely conveys the background, input, and expected output of the subtask to the model, reducing the difficulty of the model in understanding the complex problem of design text pairing. These well-designed prompts result in a significant improvement in model performance under the same task-specific training. 

For customized sentence features and sentence pair features, inspired by human reading comprehension techniques, we believe they significantly contribute to \textit{DRMiner}'s ability to extract more correct design rationales (higher recall) since they skillfully utilize latent information such as sentence structure, position, and context to assist the model in inferring challenging solutions and arguments. We will illustrate their effectiveness in RQ2 (Section \ref{subsec:rq2}).



\begin{tcolorbox}[colback=gray!10!white,colframe=gray!60!black,notitle]
Answer to RQ1: \textit{DRMiner} enhances both the quantity and completeness of the mined design rationales, particularly improving the accurate extraction of the underlying arguments.
\end{tcolorbox}

\subsection{RQ2: How effective are customized sentence features used in \textit{DSEA}?}
\label{subsec:rq2}
\begin{table}[!htbp]
\caption{The results of ablation experiments of \textit{DSEA}.}
\begin{center}
\begin{tabular}{cccc}
\toprule[1pt]
\multirow{2}{*}{\textbf{Method}} & \multicolumn{3}{c}{\textbf{Design-Related Sentence}} \\
\cmidrule{2-4}
& \textit{precision} & \textit{recall} & \textit{F1-score} \\
 \midrule
DSEA $-$ Process & 0.77 & 0.71 & 0.74 ($\downarrow 10.8\%$)\\
DSEA $-$ Position & \textbf{0.92} & 0.64 & 0.75 ($\downarrow 9.6\%$)\\
DSEA $-$ Keyword & 0.89 & 0.69 & 0.77 ($\downarrow 7.2\%$)\\
DSEA $-$ Structure & 0.90 & 0.64 & 0.75 ($\downarrow 9.6\%$)\\
DSEA $-$ Sentiment & 0.81 & 0.71 & 0.76 ($\downarrow 8.4\%$)\\
\midrule
\textit{DSEA} & 0.91 & \textbf{0.76} & \textbf{0.83} \\

\bottomrule[1pt]
\end{tabular}
\label{tab:rq3}
\end{center}
\end{table}
To show the contribution of the five dimensions of sentence features proposed in Section \ref{sub:designtextextract} to the effectiveness, we conduct five ablation experiments, each of which discarding all features in one dimension and injecting the remaining features into BERT. The results are shown in Table \ref{tab:rq3}. We find an approximately 7\%-11\% reduction in F1-scores of each experiment compared with original \textit{DSEA}, indicating that all five dimensions of features have a positive effect on the performance. 

Among them, the omission of process features has the most severe impact on the performance of \textit{DSEA}. This can be explained by the fact that this dimension of features provides subtle long-distance associations within comment area, such as relationships between the sentence author and the issue author, the total number of sentences in the comment, and the number of comments posted by the sentence author, which are hard to capture even for language models like BERT. We also observe that the precision reaches 92\% after discarding position features, which is slightly higher than original \textit{DSEA}, but there is a significant decrease in recall. We infer that sentences describing design rationales are indeed more likely to appear at specific positions in comments or comment areas (e.g., the beginning or end). However, such prior knowledge makes \textit{DSEA} more prone to categorizing unrelated sentences appearing in these positions as related, leading to a decrease in precision. This inspires us to evaluate sentences appearing at these crucial positions more thoroughly.

\begin{tcolorbox}[colback=gray!10!white,colframe=gray!60!black,notitle]
Answer to RQ2: Our customized sentence features do improve the performance of \textit{DSEA}. We believe that it is possible to continue incorporating more important features. However, it is equally critical to prevent the model from overly relying on customized knowledge.
\end{tcolorbox}

\section{Useful Evaluation for RQ3}
\label{sec:usefulexp}

Automated Program maintenance is the task of automatically fixing software defects. As highlighted in the study by Christa et al. \cite{10.1007/978-981-10-1678-3_73}, approximately 70\% of time and resources in the software industry are dedicated to maintenance activities. Moreover, more automated code generation methods are proposed and applied in practice, and there is a substantial likelihood that auto-generated code may not function as intended, particularly when addressing complex requirements. Therefore, the role of APR is becoming increasingly critical in both contemporary and future software development landscapes.


However, current researches on APR mainly focus on analyzing code itself, overlooking the crucial viewpoint that code is a result of developers' design decisions. Therefore, integrating developers' design knowledge into APR becomes a potential key point to improve the quality and efficiency of repairs. In this section, our aim is to evaluate the usefulness of the design rationales extracted by \textit{DRMiner} by providing them to advanced LLMs, e.g., \textit{CodeLLaMA-chat (CL)}, \textit{GPT-3.5-turbo}, \textit{GPT-4.0-turbo}. We compare the quality of code repairs with and without this prior design knowledge to examine its effectiveness in improving repair quality. 
For experimental data, we select 61 closed issues from \textit{Flink} tracked by Jira. We extract code snippets from the GitHub Pull Requests associated with these issues and form a dataset with 61 code fixes. The metrics we use are \textit{CodeBLEU} \cite{ren2020codebleu} and full-match number. Full-match means that the patch function generated by the model is completely consistent with the correct repair, making it more convincing than CodeBLEU. In our results, CodeBLEU is only used as a reference.

\begin{table}[!htbp]
\caption{The results of useful evaluation of design rationales extracted by \textit{DRMiner}.}
\begin{center}
\begin{tabular}{lcc}
\toprule[1pt]
\textbf{Method}  & \textbf{full-match} & \textbf{CodeBLEU} \cite{ren2020codebleu} \\
 \midrule
CL   & 1 & 0.68\\
CL $+$ DR  & 11 & 0.75\\
CL $+$ DDR  & \textbf{14} & \textbf{0.83}\\
\midrule
GPT-3.5  & 1 & \textbf{0.75}\\
GPT-3.5 $+$ DR  & 11 & 0.71\\
GPT-3.5 $+$ DDR  & \textbf{14} & 0.66\\
\midrule
GPT-4.0  & 1 & 0.75\\
GPT-4.0 $+$ DR  & 19 & \textbf{0.85}\\
GPT-4.0 $+$ DDR  & \textbf{21} & 0.84\\

\bottomrule[1pt]
\end{tabular}
\label{tab:useful}
\end{center}
\end{table}

The results are shown in Table \ref{tab:useful}. In addition to providing the design rationales (DR) extracted by \textit{DRMiner}, we also compared the situation of providing expert-annotated design rationales (ground truth, DDR). The results indicate that the performance of APR with DR is significantly improved by 14 to 19 times compared to that without, demonstrating the significant practical value of the design rationales extracted by \textit{DRMiner} in APR. We also observe that using DDR leads to approximately 11\%-27\% improvement in APR performance compared to using only DR. The reason for the difference between the two may be that DDR more accurately identifies solution and argument sentences and judges their relationships, which may be considered unrelated in \textit{DRMiner}. Thus, LLMs are able to leverage the implementation knowledge or examples contained in the design-related sentences that are misclassified as unrelated by \textit{DRMiner}. We believe that the more accurate and comprehensive the design rationales are, the greater the help they provide to APR. 

\section{Related work}
\label{sec:relatedWork}

The concept of design rationale has been proposed for over 30 years \cite{lee1991s}. Early on, some research develops systems for capturing and managing design rationales to support their reuse in software architecture, such as PAKME \cite{Babar2007TMS}, ADDSS \cite{Capilla2006WTM}, and AREL \cite{Tang2007RAM}. Although these systems can reach high effectiveness based on manually constructed knowledge repositories, the rationales usually can not be recorded concurrently with the software engineering practice due to the cost-value trade-off mentioned in Section \ref{sec:introduction}.

As people gradually discover the rich design rationale information contained in various resources \cite{Arya2019ADITO}\cite{Rana2018HDDDR}, automatically mining design rationales has attracted the interest of many researchers in recent years. Typically, \textit{Design Rationale Mining (DRM)} approaches include pattern-based, heuristic-based, and learning-based types. \textbf{Pattern-based DRM} methods \cite{Lopez2012BGS}\cite{Liang2012LWD}\cite{Burge2003RationaleSF}\cite{McCall2018} define design-related language patterns (also called ontology in a few studies), such as common vocabulary, semantic graph, or concept representation, to support recognizing and inferring over the content of rationales. \textbf{Heuristic-based DRM} methods like Rationale Miner \cite{sharma2021EROSS} and ADDRA \cite{jansen2008documenting} divide DRM into multiple steps (e.g., document selection, design recovering, rationale examination) and employ heuristic to each of them to infer the rationale behind specific decisions. There are also methods using heuristic algorithms to optimize the feature sets to improve DRM \cite{Rogers2017GDFIR}\cite{Lester2018IDRUACO}\cite{lester2020using} . 

Although pattern-based and heuristic-based DRM methods perform well on specific tasks, they struggle to adapt to the rapid growth of design-related documentation and the variability of scenarios, making it difficult to transfer them to other applications. \textbf{Learning-based DRM} methods \cite{rogers2015using}\cite{alkadhi2017RDCM}\cite{kleebaum2021continuous} can effectively mitigate this issue. They often train machine learning models with large design corpora and are guided by specific representations of text syntax and semantics to predict whether the text is related to design rationale \cite{alkadhi2017RDCM}. Besides, to integrate the advantages of above approaches, some research adopts hybrid-based DRM methods \cite{dhaouadi2022extraction}\cite{dhaouadi2022end} to achieve more accurate design rationale mining.

Additionally, several related topics bear partial resemblance to DRM. These include issue-solution pair identification \cite{Shi2021ISPY}, question and answer (Q\&A) mining \cite{Chatterjee2021AEOBQ}, design-related discussions mining \cite{viviani2019locating}, opinion mining \cite{Pang2008OMS, Bin2019PBMO}, argumentation mining \cite{Lippi2016AMS, Wyner2012SemiAutomatedAA, Palau2009AMD}, intention mining \cite{DECA, CNC}, and stance mining \cite{Parinaz2016DST, Misra2017TIIAD}. Unlike these works, design rationale encapsulates richer semantics which encompass solutions along with their supporting and opposing arguments (see Section \ref{sec:background}); thus, DRM presents a more complex challenge. Nevertheless, insights from these studies can serve as valuable guides and sources of inspiration for advancing DRM research.

\section{Conclusion and Future Work}
\label{sec:conclusions}

Open-source communities contain a wealth of design information related to software development and maintenance. However, documenting the knowledge behind these designs has not yet become a standard practice. In this paper, we propose \textit{DRMiner} to automatically extract latent design rationales from issue logs. The key idea behind \textit{DRMiner} is to ({\romannumeral1}) use ingenious prompt to improve the performance of language models, and ({\romannumeral2}) inject customized sentence features to reveal the latent relationships within comments. Experimental results show that \textit{DRMiner} surpasses existing methods in both quality and quantity of extracted design rationales. Moreover, we demonstrate that the extracted design rationales significantly improve the quality of automated program repair. In the future, we will explore integrating design knowledge across communities (e.g., GitHub, Gitee, Sourceforge, OSChina) to build a dynamically updated design rationale repository. We will also explore the effectiveness of the automatically mined design rationales on a larger scale of code defects.

\newpage
\balance
\bibliographystyle{IEEEtran}
\bibliography{reference}

\end{document}